\documentclass[acmsmall,screen,authorversion]{acmart}
\usepackage{tabulary}
\usepackage{hyperref}

\AtBeginDocument{%
  }

\setcopyright{acmcopyright}
\copyrightyear{2023}
\acmYear{2023}

\acmConference[GenAICHI Workshop]{Generative AI and HCI Workshop}{April 28 2023}{Hamburg, Germany}
\acmBooktitle{CHI '23 Worksop on Generative AI and HCI, April 23--28, 2023, Hamburg, Germany}
\acmPrice{15.00}
\acmISBN{XXX-X-XXXX-XXXX-X/XX/XX}

\makeatletter
\def\@copyrightspace{\relax}
\makeatother

\begin{document}

\title[Generative AI for News Media]{Envisioning the Applications and Implications of Generative AI for News Media}

\author{Sachita Nishal}

\author{Nicholas Diakopoulos}
\affiliation{%
  \institution{Northwestern University}
  \country{USA}
}

\renewcommand{\shortauthors}{Nishal and Diakopoulos}



\maketitle

\section{Introduction}
In January 2023,  \textit{CNET} published algorithmically generated news articles and financial advice columns under the ambiguous byline of "CNET Money" \cite{landymoreCNETQuietlyPublishing2023}, with a small disclaimer offered beneath: "This article was assisted by an AI engine and reviewed, fact-checked and edited by our editorial staff." \cite{cnetstaffCNETMoney2022}. In February 2023, Men's Journal (which has the same publisher as \textit{Sports Illustrated}) followed suit, with their AI-generated, human-edited articles carrying the following disclaimer: "This article is a curation of expert advice from Men’s Fitness, using deep-learning tools for retrieval combined with OpenAI’s large language model for various stages of the workflow." \cite{bruellSportsIllustratedPublisher2023}. 

Algorithmically generated news articles have been published for almost a decade in finance, sports, weather, and in other domains where structured data is available  \cite{diakopoulosAutomatingNewsHow2019}. 
The key distinction between earlier deployments and what is happening at \textit{CNET} or \textit{Men's Journal} now is that models are no longer using straightforward template-based approaches for text generation from structured data. Instead, generative AI is being used to write more substantial drafts of articles that purport to offer advice, discuss subjective topic ideas, or in other cases, such as BuzzFeed, personalize interactive content for readers \cite{lundenBuzzFeedLaunchesInfinity2023}. But these examples also highlight the risks of deploying generative models in journalism, in particular since articles from both \textit{CNET} and \textit{Men's Journal} were riddled with factual errors despite the publishers claiming human review \cite{christianCNETArticleWritingAI2023, christianMagazinePublishesSerious2023}. Beyond accuracy, generative AI may have unanticipated impacts: at \textit{CNET} the system generated stories rapidly, but editing took more time than editing a human journalist’s work \cite{satoCNETPushedReporters2023}.


Generative AI performance continues to improve across tasks such as abstractive summarization, audio transcription, machine translation, and so on \cite{brownLanguageModelsAre2020, radfordRobustSpeechRecognition2022}. Due to these capabilities, they present promise for augmenting journalists' work beyond helping them generate article drafts, such as for simplifying newsworthy but highly technical documents [under review], illustrating articles visually \cite{liuOpalMultimodalImage2022, newmanJournalismMediaTechnology2023}, or as a creativity support tools \cite{petridisAngleKindlingSupportingJournalistic2023}. Technologies incorporating more traditional approaches in supervised learning and natural language processing have been successfully piloted and used in newsrooms to support a slew of tasks across the news production pipeline \cite{kobieReutersTakingBig2018, magnussonFindingNewsLead2016, diakopoulosAutomatingNewsHow2019, liuReutersTracerAutomated2017}. In this workshop paper, we explore how generative AI can similarly be incorporated into hybrid task workflows of human-AI collaboration in newsrooms, while articulating the ethical and journalistic values that must be negotiated in designing and developing the technology and interfaces to use it effectively in the domain. 

In this paper we explore the following high-level questions integrating research in computational journalism and HCI \cite{aitamurtoHCIAccurateImpartial2019, diakopoulosAutomatingNewsHow2019, broussardArtificialIntelligenceJournalism2019, simonUneasyBedfellowsAI2022, gutierrezlopezQuestionDesignStrategies2022, dehaanInvisibleFriendFoe2022}: which tasks within the news production pipeline might benefit from generative support? What would be the repercussions be for journalistic agency and creativity? What are editorial and journalistic values that should influence the design of these systems? What modes of interaction might support human-AI collaboration in news production? The following sections explore these questions and suggest beneficial areas for future work. 


\section{The Suitability of Generative AI for Newsroom Tasks}

Here we map out a few journalistic tasks and consider which  might benefit from the use of generative AI. We also expound on the journalistic values that must be considered when designing this technology and interfaces to it.

A recent Associate Press report synthesized insights about the deployment and workflows of AI and related technologies in over a hundred American newsrooms \cite{rinehartArtificialIntelligenceLocal2022a}. AP interviewed editors, journalists, media executives, and marketing managers to understand how they leveraged or experimented with AI while producing news stories, and what innovations could be useful. Broadly speaking, there was a preference for (i) human-supervised systems (ii) that would be sensitive to the confidentiality and privacy of their sources, (iii) and would have "low cost, low learning curve, and low maintenance" \cite{rinehartArtificialIntelligenceLocal2022a}.  

In response to these needs, we note that generative AI lends itself well to an interactive paradigm involving human instruction and feedback \cite{duReadReviseRepeat2022, ouyangTrainingLanguageModels2022}. However, text generation models have been shown to memorize and reveal sensitive information from training data \cite{carliniExtractingTrainingData2021}, and OpenAI's policy for their models even explicitly states that prompts input by users could be used to improve model performance in the future \cite{openaistaffChatGPTGeneralFAQ2023}. This could hinder basic applications such as text summarization if journalists prompt ChatGPT using confidential data or documents \cite{eliotGenerativeAIChatGPT2023}. Generative AI could also become expensive due to extensive experimentation as users trial different prompts to elicit their desired output \cite{pruthiCombatingAdversarialMisspellings2019, moradiEvaluatingRobustnessNeural2021}, which could limit use in contemporary newsrooms as they already struggle with revenues and operating costs \cite{matsaLocalNewspapersFact2022, barthelKeyTakeawaysState2021}. Cost is crucial for smaller newsrooms to reap benefits of generative AI, since they often lack dedicated technologists needed to build and maintain such tools.
Finally, we recognize that broader initiatives to build basic AI competencies (e.g. prompt engineering) are necessary to ensure a low learning curve for journalists \cite{longWhatAILiteracy2020, deuzeImaginationAlgorithmsNews2022}.

Beyond these general insights, four main sub-areas of tasks where automation could potentially be used emerged from the AP's interviews: newsgathering, production, distribution, and business. In this short paper we center journalists and editors and focus on applications for \textit{newsgathering} and \textit{production} in particular.

\subsection{Newsgathering}

Newsgathering relates to the sourcing and investigation of potentially newsworthy leads, before a journalist or editor decides to pursue their development into a full-fledged story \cite{mcmanusMarketdrivenJournalismLet1994}. This entails the initial discovery of a story from sources such as a journalist's personal network, press releases, administrative documents, and so on. Journalists then engage in a process of vetting and sense-making, as they try and ascertain if the initial lead lends itself to development into a news item \cite{reichProcessModelNews2006}. Recent algorithmic support systems for newsgathering have supported content discovery via statistical evaluations of "newsworthiness" for leads, as well as via automated detection of anomalous phenomena \cite{diakopoulosUnderstandingSupportingJournalistic2021, nishalCrowdRatingsPredictive2022, magnussonFindingNewsLead2016, diakopoulosGeneratingLocationBasedNews2020, liuReutersTracerAutomated2017, wangJournalisticSourceDiscovery2021}. Systems to support further sense-making, based on the potential framing and narratives around a lead have also been proposed \cite{franksUsingComputationalTools2021}. Similar to this prior work, we do not argue that generative AI can substitute existing routines of news discovery and sense-making, but we do believe it can supplement them in the ways described below. 

\subsubsection{Applications of Generative AI: Summarization and Querying}

Interviewees quoted in the AP report expressed a desire for AI interventions to support content discovery from both structured datasets (e.g. court records, police records) and unstructured datasets (e.g. legislative documents, consumer reports). Given the below-average statistical and mathematical capabilities of LLMs \cite{friederMathematicalCapabilitiesChatGPT2023}, we propose that these models mainly be used to supplement content discovery and sense-making from unstructured datasets and text documents.

To support more rapid discovery and evaluation of newsworthy information, LLMs can be used for extractive and abstractive summarization of unstructured texts \cite{nallapatiSummaRuNNerRecurrentNeural2017, paulusDeepReinforcedModel2018}, especially for complex and jargon-heavy documents. 
These activities can be executed within interfaces that (i) \textbf{provide pre-computed text generations}, either as an initial summary, or selected from a pre-made set of potential prompts and their outputs, with prompts engineered after extensive experimentation by technical experts, e.g. prompts to show a brief summary, or "newsworthy" angles for the given text, or retrieve past news coverage on similar topics as the text, or (ii) \textbf{support querying and brainstorming} with an LLM via means of chatbot-style interactions, to ask specific questions about the document at hand, requesting quotes, clarifications, and even counter-arguments to its claims, as necessary. Recent advances in using LLMs to support ideation during a reporter's initial encounter with a lead have shown promise \cite{petridisAngleKindlingSupportingJournalistic2023}, and even point to the benefits of personalizing LLM  outputs to better align with journalistic interests.

\subsubsection{Tensions with Journalistic Values and Norms}

One consideration is that abstractive summarization can yield "hallucinations": text generations that do not necessarily adhere to the information presented in the input prompt, and could be factually incorrect. Further, authoritative and confident presentation of an LLM's responses can lead to over-reliance and unwarranted trust, especially when end-users are unfamiliar with the original text, or if the model generates citations \cite{chengMappingDesignSpace2022, howe2022exploring, duReadReviseRepeat2022}. This can threaten the journalist's goal of reporting accurate and credible information \cite{hanitzschDeconstructingJournalismCulture2007}, especially in light of how fact-checking and verification in the newsroom are often conducted with pragmatic compromises and sometimes with a default reliance on sources \cite{barnoyWhenWhyHow2019, diekerhofCheckNotCheck2012, vanwitsenHowScienceJournalists2021}. 

While technical work in remedying hallucinations is underway \cite{liGuidingGenerationAbstractive2018, jiSurveyHallucinationNatural2022a}, user studies of how reporters engage with the proposed applications of generative AI are equally vital: to what extent and in what ways do journalists scrutinize an LLM's outputs at the initial stage of content discovery and sense-making? What is the perceived risk of hallucinations? What kinds of textual disclaimers can decrease (or increase) journalists' reliance on outputs? Do visual or graphical representations of uncertainty in outputs influence trust? How do journalists make trade-offs between their limited time availability, and the relative freedom of experimentation provided by some interaction paradigms (e.g. querying) over others (e.g. pre-computed summaries)? How do instances of model bias at this stage interact with journalistic objectivity \cite{deuzeWhatJournalismProfessional2005}? Aligning generative AI for use in the domain will benefit from grappling with these questions. 



Another aspect is that journalistic assessments of "newsworthiness" at this stage involve highly contextual decision-making around whether a story exhibits certain \textit{news values}, i.e. if it is controversial, or surprising, or novel, etc., for a reporter's intended audience \cite{harcupWhatNewsNews2017}. Engineering prompts to explicitly summarize and query text through these lens is important so that the results are aligned with journalistic values. Conducting human evaluation studies with the designed prompts across various news values and specific beats (e.g. science, law, policy), could also help identify the context-specific abilities of generative AI.

\subsection{News Production}

News production relates to activities involved in developing a story, including the iterative processes of writing and editing, and the creation of material for advertising and distribution. During this process, journalists will select reporting formats, center certain news values, interview sources, receive editorial feedback, and so on. A range of individual, structural, and cultural factors influence these activities: the self-perceived roles of journalists (e.g watch-dogging, dissemination), their normative ideologies, and the market orientation of the news organization \cite{fahyScienceJournalistOnline2011, donsbachJournalistsRolePerception2012, hanitzschDeconstructingJournalismCulture2007, allernJournalisticCommercialNews2002}. Algorithmic support systems for writing are often aimed at a broad audience (e.g. Grammarly,  Hemingway) and typically provide spell checks, grammatical fixes, stylistic feedback, and readability assessments - simpler features that do not require creative, argumentative, or factual input from the support system. Here we reflect on how generative AI could support highly specialized tasks involved within news production. 

\subsubsection{Applications of Generative AI: Collaborative Writing and Content Creation}

Interviewees surveyed within the AP report expressed a desire for automated writing support for both structured and unstructured data, as well as for social media content creation. We reiterate that writing with generative AI from structured datasets is error-prone, but \textbf{summarization of unstructured documents} could help kickstart a journalist's writing process. This would require a journalist to craft their prompt to specify the desired features of the summary (e.g. short, or bulleted, or with a higher-level of simplicity), and would still require human oversight to ensure factuality. This process would likely augment some reporting formats (e.g. short summaries of sporting events) more than others (e.g. long-form features). Generative AI can further be used to \textbf{iteratively propose edits for human-written text}, with explanations \cite{reidLearningModelEditing2022, chuaHowChatbotsCan2023}. The system could suggest edits to improve fluency or coherence, and also in response to specific human instructions or plans \cite{duReadReviseRepeat2022, schickPEERCollaborativeLanguage2022}. Summaries of a news item for social media channels and publicity can also be generated.

\subsection{Tensions with Journalistic Values and Norms}

Interviews with journalists indicate a general optimism that automated writing for mundane, repetitive tasks can free them up for in-depth reporting "requiring the skills that human journalists embody" \cite{schapalsAssistanceResistanceEvaluating2020}. This points to another latent tension in automated writing and the journalistic process: that machines cannot always augment deeper journalistic intuitions, worldviews, and values  \cite{bucherMachinesDonHave2017} and could both support or undermine specific journalistic views and values when used for more complex writing \cite{komatsuAIShouldEmbody2020}. Interviewees considered how suggestions from AI-based writing systems could help them be impartial by suggesting new perspectives, or impose biases from the training data in their writing. Similarly, journalists believed that generative AI could either support originality by sparking fresh ideas, or constrain it via cliched recommendations or by broadly devaluing reporters' hunches. 

The duality of potential impact suggests that interface design for writing support should offer explanations for suggestions, so reporters can understand why edits are suggested, and how they interact with their own value priorities \cite{abdulTrendsTrajectoriesExplainable2018a}. Designing writing support systems in partnership with journalists to proactively support specific values can also be tested, e.g. Komatsu et al. suggest that fine-tuning models to detect and indicate hyper-partisan orientation in source documents or writing can support goals such as objectivity and impartiality \cite{komatsuAIShouldEmbody2020, deuzeWhatJournalismProfessional2005}. However, this could undermine efficiency by introducing extensive prompt engineering and writing evaluation into the workflow. Given that journalists could be hesitant to reject the outputs of generative models \cite{komatsuAIShouldEmbody2020, howe2022exploring}, interfaces could also benefit from representations of uncertainty around edit suggestions, or even operationalizing  thresholds on the probability of suggested edits \cite{alammarEccoOpenSource2021}.

Ultimately, it is worth reiterating that journalistic writing serves a wide variety of roles, speaks to differing contexts and news values, and highly depends on a journalist's own voice and intent. This makes designing or evaluating fine-tuned models, specialized interfaces, user studies, and evaluation metrics more difficult as compared to another common case of AI-augmented content generation - writing code \cite{zieglerProductivityAssessmentNeural2022a, mozannarReadingLinesModeling2022}. Journalism studies reminds us time and again that the newsworthiness of events is not inherent, but rather generated and articulated by journalists based on the surrounding context \cite{lesterGeneratingNewsworthinessInterpretive1980, bucherMachinesDonHave2017}. Realistically, this means that interfaces for writing support systems must be designed to enable greater agency and transparency for journalists, so that they are well-informed and in greater control as they receive support from generative AI for the kind of stories they want to tell.

\section{Conclusion}

In this paper, we have explored how generative models can help reporters and editors across a range of tasks from the conception of a news story to its distribution. All these applications would be enacted within the complex socio-technical system of journalists, editors, executives, and so on, meaning that some tensions that consequently arise will inherently be personal, political or structural \cite{mokanderAuditingLargeLanguage2023}. We have outlined some of these tensions here, and we stress that this necessitates inclusive, value-sensitive design processes for the use and evaluation of generative AI in the newsroom, conducted in partnership with journalists, editors and other participants.

\bibliographystyle{ACM-Reference-Format}
\bibliography{main-bib}

\end{document}